# Is the quantum hydrodynamic analogy more general than the Schrödinger approach?

Piero Chiarelli

*National Council of Research of Italy, Area of Pisa, 56124 Pisa, Moruzzi 1, Italy Interdepartmental Center*
*"E.Piaggio" University of Pisa*
Phone: +39-050-315-2359
Fax: +39-050-315-2166

Email: pchiare@ifc.cnr.it.

**Abstract:** The quantum hydrodynamic analogy (QHA) equivalent to the Schrödinger equation is investigated and extended to the stochastic case. The investigation shows that in addition to reproducing the standard quantum mechanics the QHA model is able to generally describe the quantum stochastic dynamics leading to the dissipative Schrödinger equation given by Kostin [55] as a particular case. The inspection shows that the QHA is well suited for the treatment of problems where local noise (spatially distributed one) is introduced. In this case the analysis shows that the bi-univocal correspondence between the QHA and the Schrödinger approach breaks down and that the states described by the QHA do not have their corresponding ones into the Schrödinger description.

## 1. Introduction

Since the introduction by Schrödinger of the quantum wave equation, it was presented by Madelung [1] an equivalent approach to quantum mechanics that gives rise to an interesting logical approach to the quantum dynamics.
The interest for the quantum hydrodynamic analogy (QHA) of quantum mechanics had never interrupted since nowadays. It has been studied and extended by many authors as Jánossi [2] resulting useful in the numerical solution of the time-dependent Schrödinger equation [3-5].
More recently it has been used for modeling quantum dispersive phenomena in semiconductors that cannot be described by the semi-classical approximation [6,7].
As put in evidence by Tsekov [8-9], compared to others classical-like approaches (e.g., the stochastic quantization procedure of Nelson [10-12], the mechanics given by Bohm [13-16] and other ones proposed by Takabayasi [17], Guerra and Ruggiero [18], Parisi and Wu [19-20]) and others [21], the QHA has the precious property to be exactly equivalent to the Schrödinger equation (giving rise to the same results) and is free from problems such as the undefined Bohmian variables [8-9] or the confusion between the statistical and the quantum fluctuations as in the Nelson theory [10-12].
The Schrödinger equation is by far more manageable for calculation than the QHA while it owns an idiosyncrasy with the local character of large scale problem. On the other hand, the QHA is practically intractable for any physical problem but it owns a classical-like structure that makes it suitable for the achievement of a overall understanding of quantum and classical phenomena. The classical like-approach has resulted useful in explaining phenomena such as the dispersive effects in semiconductors [6,7] multiple tunneling [22], mesocopic and quantum Brownian oscillators [23], critical phenomena [24-26], and theoretical regularization procedure of quantum field [27-28].
Here we show the physical relevance of the QHA goes beyond the equivalence with the Schrödinger problem and that is able to reproduce among which, as a particular case, the Schrödinger-Langevin equation given by Kostin. Moreover, we also show that it possibly can appropriately treat the introduction of local noise (spatial distributed one) into the quantum stochastic problem.

## 2. The quantum hydrodynamic analogy

The QHA-equations are based on the fact that the Schrödinger equation, applied to a wave function $\psi_{(q,t)} = A_{(q,t)} \exp[i\, S_{(q,t)}/\hbar]$, is equivalent to the motion of a fluid with particle density $n_{(q,t)} = A^2_{(q,t)}$ and a velocity $\dot{q}_\alpha(q,t) = m^{-1} \partial S_{(q,t)}/\partial q_\alpha$, governed by the equations

$$\partial n_{(q,t)}/\partial t + \partial (n_{(q,t)} \dot{q}_\alpha)/\partial q_\alpha = 0, \tag{1}$$

$$(\dot{q}_\alpha, \dot{p}_\beta) = (\partial H/\partial p_\alpha, -\partial(H+V^{qu})/\partial q_\beta) \equiv \upsilon^Q_j, \qquad 1 \leq \alpha, \beta \leq 3n, \qquad 1 \leq j \leq 6n \tag{2}$$

where Greek indices are used for ordinary $3n$-vectors while alphabetic indices are used for phase-space $6n$-vectors (i.e., for $j \leq 3n \Rightarrow \upsilon^Q_{j=\alpha} = \partial H/\partial p_\alpha$ and for $j > 3n \Rightarrow \upsilon^Q_{j=3n+\beta} = -\partial(H+V^{qu})/\partial q_\beta$). By defining $\upsilon^H_j \equiv (\partial H/\partial p_\alpha, -\partial H/\partial q_\beta)$, $\upsilon^{qu}_j \equiv (0, -\partial V^{qu}/\partial q_\beta)$ we can ideally subdivide the phase-space velocity into the Hamiltonian and quantum ones to read $\upsilon^Q_j = \upsilon^H_j + \upsilon^{qu}_j$. Moreover, $n$ is the number of structureless particles of the system whose mass is $m$ and $V^{qu}$ is the quantum pseudo-potential that originates the quantum non-local dynamics and reads

$$V^{qu}_{(q,t)} = -(\hbar^2/2m)\, n_{(q,t)}^{-\frac{1}{2}}\, \partial^2 n_{(q,t)}^{\frac{1}{2}}/\partial q_\alpha \partial q_\alpha. \tag{3}$$

### 2.1. The quantum hydrodynamic analogy as probability density function of a Stochastic motion equation

For the purpose of this paper, it is useful to observe that equations (1-2) can be derived by the following equations

$$\partial \rho/\partial t + \partial\, \upsilon^Q_j \rho/\partial x_j = 0, \tag{4}$$

where $x_j \equiv (q_\alpha, p_\beta)$, once equation (4) is integrated over the momentum $p_\alpha$ (i.e., $\int_{-\infty}^{+\infty} \ldots \int_{-\infty}^{+\infty} dp_1 \ldots dp_{3n}$) with the sufficiently general condition that

$$\lim_{|p| \to \infty} \rho_{(q,p,t)} = 0 \tag{5}$$

and the PDF "$\rho$" has the form

$$\rho_{(q,p,t)} = n_{(q,t)} \Pi_\alpha \delta(p_\alpha - \partial S_{(q,t)}/\partial q_\alpha), \tag{6}$$

with

$$n_{(q,t)} = \int_{-\infty}^{+\infty} \rho_{(q,p,t)}\, dp_1 \ldots dp_{3n}. \tag{7}$$

Moreover, let's consider the stochastic differential equation

$$\dot{x}_j\, dt = \upsilon^Q_j\, dt + (k\Theta)^{\frac{1}{2}} \mu^{\frac{1}{2}}_{jm}\, d\mathcal{W}_m, \tag{8}$$

so that (4) can be seen as the deterministic limit of a Markovian process where k is the Boltzmann constant, $\Theta$ is the fluctuation amplitude parameter, $\mu_{mj} = \mu^{\frac{1}{2}}_{mn} \mu^{\frac{1}{2}}_{nj}$ is the associated migration tensor and $d\mathcal{W}_m$ is a white noise with null mean and unitary variance [49,52].
In the quantum case, the pseudo-potential $\upsilon^{qu}_j$ is a function of the probability density function (PDF) $\rho$ and some differences arise with respect to the classical case. These clearly appear if we integrate the stochastic differential equation (SDE) (8) by means of the Cauchy-Euler method.

Looking at the solution for the probability transition functions (PTF) of (8) [49, 53-54] that reads

$$P(x_{(t)}, x_0 | t, 0) = \lim_{\Delta t \to 0} \left\{ \prod_{n=1}^{w=(t/\Delta t)-1} \int_{-\infty}^{+\infty} dx_n \right\} P(x_{n+1}, x_n | (n+1)\Delta t, n\Delta t) \, P(x_1, x_0 | \Delta t, 0), \quad (9.a)$$

with

$$P(x_{n+1}, x_n | (n+1)\Delta t, n\Delta t) \propto \exp{-\tfrac{1}{2}}[(x_{j(n+1)} - \langle x_{j(n+1)} \rangle)(k\Theta\Delta t)^{-1}\mu^{-1}_{jm}(x_{m(n+1)} - \langle x_{m(n+1)} \rangle)]$$

$$\propto \exp{-\tfrac{1}{2}}[(x_{j(n+1)} - (x_{j(n)} + \upsilon^Q_{j(n)}\Delta t))(k\Theta\Delta t)^{-1}\mu^{-1}_{jm}(x_{m(n+1)} - (x_{m(n)} + \upsilon^Q_{m(n)}\Delta t))]. \quad (9.b)$$

we can see that, on the contrary of the classical case, $\upsilon^Q$ is not a defined function of the phase space but changes at each step of integration since $\upsilon^Q_{(x_n, (n+1)\Delta t)}$ through (3) is defined by $\rho_{(x_{n+1},(n+1)\Delta t)}$ that is defined by $\upsilon^Q_{(x_n, n\Delta t)}$ and $\rho_{(x_n, n\Delta t)}$ at previous time (and it can be calculated by using the conservation of $\rho$).

Since (9.a) obeys to the Smoluchowski integro-differential equation [52], it is possible to derive the differential conservation equation for the PTF P(x,z|t,0) by transforming the Smoluchowski equation into a differential one by the method of Pontryagin [52] to get

$$\partial P_{(x,z|t,0)} / \partial t + \partial P_{(x,z|t,0)} \upsilon_j / \partial x_j = 0, \quad (10)$$

where the current $P_{(x,z|t,0)} \upsilon_j$ reads

$$P_{(x,z|t,0)} \upsilon_j = P_{(x,z|t,0)} \upsilon^Q_j - \tfrac{1}{2} \partial \mathbf{b}_{jm} P_{(x,z|t,0)} / \partial x_m + \sum_{n=2}^{\infty} (n!)^{-1} \partial^n c^{(n)}_{jk\ldots m} P_{(x,z|t,0)} / \partial x_k \ldots \partial x_m, \quad (11)$$

with

$$c^{(n)}_{jk\ldots m\,\{n+1\text{indexes}\}} = \lim_{\tau \to 0} \tau^{-1} \int_{-\infty}^{+\infty} (y_j - x_j)(y_k - x_k) \ldots (y_m - x_m) P(y,x|\tau,t) \, dy_k \ldots dy_m. \quad (12)$$

Usually, in the Hamiltonian (i.e., classical) case, the distance $(y_k - x_k)$ (attainable with a certain probability $P(y,x|\tau,t)$) becomes smaller and smaller as $\tau$ shorter and shorter so that: (1) Velocities $\lim_{\tau \to 0} (y_k - x_k)/\tau$ are finite. (2) The squared mean displacements are proportional to $\tau$ (i.e., $\lim_{\tau \to 0} \langle (y_k - x_k)^2 / \tau \rangle \equiv \mathbf{b}_{kk}$ finite). (3) The cumulants $c^{(n)}_{jk\ldots h}$ go like $\tau^{n-2}$ and, hence, since $\tau$ can be chosen arbitrarily small, $c^{(n)}_{jk\ldots h} \to 0$ for $n > 2$ [52] and the differential equation (10) becomes a FPE with Gaussian PTF.

On the contrary, the presence of the quantum potential gives a functional dependence of $\upsilon^Q$ by the PDF $\rho_{(x, t)}$, so that equation (10) is not a FPE, the PTF $P(y,x|\tau,t)$ (9.a) is not Gaussian [54] and cumulants higher than two are not null.

## 2.2. The quantum hydrodynamic analogy in the stochastic case

By posing $t = 0$, $\tau = t$ and integrating (10) such as $\rho(x,t) = \int_{-\infty}^{+\infty} P(x,x_0|t,0) \, d^{6n}x_0$ it follows that

$$\partial \rho / \partial t + \partial \rho \upsilon_j / \partial x_j = 0 \quad (13)$$

$$\rho \upsilon_j = \rho \upsilon^Q_j - \tfrac{1}{2} \partial \mathbf{b}_{jm} \rho / \partial x_m + \ldots + \sum_{n=2}^{\infty} (n!)^{-1} \partial^n c^{(n)}_{jk\ldots m} \rho / \partial x_k \ldots \partial x_m. \quad (14)$$

To obtain the QHA-like representation of the partial differential equation (PDE) (13) for $\Theta \neq 0$, it is sufficient to integrate it over $p_\alpha$ with the condition that $\lim_{|p| \to \infty} P(q,p,t) = 0$. Therefore, with the help of (2, 3, 14) we obtain

$$\partial n_{(q,t)}/\partial t + \partial (n_{(q,t)} \bar{\dot{q}}_{\alpha(q,t)})/\partial q_\alpha = 0, \tag{15.a}$$

$$\dot{q}_\alpha = \partial H/\partial p_\alpha - (2\rho)^{-1}[\partial (\mathbf{B}_{\alpha\beta}\,\rho)/\partial q_\beta + \partial (\mathbf{B}_{\alpha\,(3n+\beta)}\,\rho)/\partial p_\beta], \tag{15.b}$$

$$\dot{p}_\alpha = -\partial (H+V^{qu})/\partial q_\alpha - (2\rho)^{-1}[\partial (\mathbf{B}_{(3n+\alpha)\beta}\,\rho)/\partial q_\beta + \partial (\mathbf{B}_{(3n+\alpha)(3n+\beta)}\,\rho)/\partial p_\beta], \tag{15.c}$$

where

$$\mathbf{B}_{jm} = -\mathbf{b}_{jm} + 2(\rho\,n!)^{-1} \sum_{n=2}^{\infty} \partial^{n-1} \mathbf{c}^{(n)}_{j\,k\ldots m-1,\,m}\,\rho/\partial x_k \ldots \partial x_{m-1}, \tag{16.a}$$

$$\bar{\dot{q}}_\alpha = \int_{-\infty}^{+\infty} \dot{q}_\alpha\,\rho_{(q,p,t)}\,dp_1\ldots dp_{3n}\,/\,n_{(q,t)}. \tag{16.b}$$

### 2.3. Deterministic limit: the quantum mechanics

It is interesting to see the conditions under which the QHA-equations (1-2) can be derived by equations (15.a-15.c). In the deterministic limit, equations (13-14) reduces to equations (4-5) with (9.a) that reads

$$\lim_{\Theta \to 0} P(y,x|\tau,t) = \delta(y-(x + \upsilon^Q \tau)), \tag{17}$$

By observing that the ensemble of solutions of equations (4-5) obeying to conditions (17) is wider than that one of the QHA-equations (1-2), it comes clearly out that the accessory condition (6) must be imposed to (13-14) (in the deterministic limit) in order that equation (15.a) leads to (1).
By inserting (6) into (16.b) we see that the condition required reads

$$\bar{\dot{q}}_\alpha = \dot{q}_\alpha = m^{-1}\,\partial S_{(q,t)}/\partial q_\alpha. \tag{18}$$

(namely the wave-particle equivalence). It must be underlined that condition (18) is necessary to pass back from the equations (15.a-15.c) to the Schrödinger representation, since, generally speaking, $\bar{\dot{q}}_\alpha \neq m^{-1}\,\partial S_{(q,t)}/\partial q_\alpha$.

Therefore, as shown in appendix A, in order to warrant that equation (13) satisfies the wave-particle equivalence in the quantum deterministic limit, for system whose Hamiltonian has the form $H(q,p) = p_\alpha p_\alpha/2m + V(\mathbf{q})$, we have to reduce the noise to spatial-momentum decoupled one of type

$$(k\,\Theta)^{1/2}\,\mu^{1/2}_{jm} = D^{1/2}_{jm} = \begin{pmatrix} 0 & 0 \\ 0 & D^{1/2}_{\mathbf{p}\,\alpha\beta} \end{pmatrix} \tag{19}$$

It is worth noting that in the deterministic limit the PDF $\delta$-peaked shape (6) as well as (18) are warranted along time if at a certain initial time $t_0$, the PDF has the form $\rho_{(q_O,\,p_O)} = n_{(q_O)}\,\Pi_\alpha \delta(p_{O\alpha} - \partial S/\partial q_\alpha(q_O))$. In fact, by introducing (17) into the integral conservation equation for the PDF, at a generic instant $t_0 + t$, it follows that

$$\rho_{(q(t_0+t),p(t_0+t))} = \int_{-\infty}^{+\infty} n_{(q(t))}\,\Pi_\alpha \delta(p_{\alpha(t_0)} - \partial S/\partial q_\alpha(q(t_0)))\,\delta(q_{(t_0)}-q_{(t_0+t)})\delta(p_{(t_0)}-p_{(t_0+t)})d^{3n}q\,d^{3n}p = n_{(q(t_0+t))}\,\Pi_\alpha \delta(p_{\alpha(t_0+t)} - \partial S/\partial q_\alpha(q_{(t_0+\tau)})).$$

Hence, among the deterministic states of (2,4) (i.e., of the standard quantum mechanics) the condition (6) is self-sustained and it holds forever if owned at a certain initial time.

### 2.4. The Brownian harmonic oscillator

When the TPF has the form $\lim_{\tau \to 0} P(y,x|\tau,t) = \delta(y-(x + \upsilon\,\tau))$ and the PDF owns the form (6), the system of equations (15.a-15.c) converges to (1-2) and it can be always traced back to the Schrödinger representation thanks to the necessary relation (18) that in this case holds.

More generally, given the deterministic QHA with a Hamiltonian $H=H_0+V_{e\,(q,p,t)}$ where $H_0$ is the system's Hamiltonian and where $V_{e\,(q,p,t)}$ describes an additional interaction that may also contain the external environment, the Schrödinger equation can be generally written in term of a Hamiltonian $H=H_0+V_{e(q,\partial S/\partial q,t)}$ where

$$V_{e\,(q,\partial S/\partial q,t)} = \int_{-\infty}^{+\infty} V_{e\,(q,p,t)}\, \Pi_\alpha \delta[p_\alpha - (\partial S_{(q,t)}/\partial q_\alpha)]\, d^{3n}p_\alpha. \qquad (20)$$

When the external environment is present through $V_{e(q,\partial S/\partial q,t)}$, the QHA may comprehend some stochastic dynamics. In fact, if it is possible to define a complex "stochastic potential" (SP) $G_{ik}$ such as

$$\upsilon^{SP}_j = \upsilon_j - \upsilon^Q_j = -\tfrac{1}{2}(\rho)^{-1}\partial\, \mathbf{b}_{jm}\,\rho/\partial x_m + \ldots + (\rho\, n!)^{-1} \sum_{n=2} \partial^n\, \mathbf{c}^{(n)}_{jk\ldots m}\,\rho/\partial x_k\ldots\partial x_m$$

$$= (\partial \mathrm{Im}[G_{\alpha\beta}]/\partial p_\beta,\, -\partial \mathrm{Re}[G_{\alpha\beta}]/\partial q_\beta). \qquad (21)$$

equation (13) reads

$$\partial\rho/\partial t + \partial(\upsilon^Q+\upsilon^{SP})_j\,\rho\,/\partial x_j \qquad (22)$$

that, analogously to (4) with (15.a) leads to

$$\partial n_{(q,t)}/\partial t + \partial(n_{(q,t)}\,\dot{q}_\alpha\,(q,t))/\partial q_k = 0, \qquad (23)$$

$$\dot{q}_\alpha = \partial(H\,\delta_{\alpha\beta}+ \mathrm{Im}[G_{\alpha\beta}])/\partial p_\beta, \qquad (24)$$

$$\dot{p}_\alpha = -\partial(H\,\delta_{\alpha\beta}+ V^{qu}\,\delta_{\alpha\beta}+ \mathrm{Re}[G_{\alpha\beta}])/\partial q_\beta. \qquad (25)$$

In order (23 - 25) to be equivalent to (2,4), in the case of identical structureless particles of mass $m$, for the Hamiltonian of type $H_{0(q,p)} = p_\alpha p_\alpha/2m + V(\mathbf{q})$, it must hold

$$\partial V_{e\,(q,p,t)}/\partial q_\alpha = (\partial \mathrm{Re}[G_{\alpha\beta}]/\partial q_\beta) - 2m\, d(\partial \mathrm{Im}[G_{\alpha\beta}]/\partial p_\beta)/dt. \qquad (26)$$

An interesting example of dynamics described by (20- 25) is the Brownian harmonic oscillator (BHO) that is obtained by posing

$$\mathrm{Im}[G_{\alpha\beta}] = 0, \qquad (27)$$

$$\partial \mathrm{Re}[G_{\alpha\beta}]/\partial q_\beta = [\beta(p_\alpha/m) - F_\alpha(t)], \qquad (28)$$

where $F_\alpha(t)$ is the stochastic force and $\beta$ is the friction coefficient to which the particle is submitted as defined in Ref. [64, 66].
Hence, the BHO phase space velocity reads

$$\upsilon^{BHO}_\beta = \upsilon^Q_\beta + (\partial \mathrm{Im}[G_{\alpha\beta}]/\partial q_\beta,\, \partial \mathrm{Re}[G_{\alpha\beta}]/\partial q_\beta) = \upsilon^Q_\beta + (0, [\beta(p_\alpha/m)-F_\alpha(t)]) \qquad (29)$$

By introducing conditions (6, 27 – 28) in (26) it is possible to calculate the system-bath interaction potential $V_{e(q,p,t)}$ for the BHO in the hydrodynamic representation that reads

$$\partial V_{e\,(q,p,t)}/\partial q_\alpha = [\beta(p_\alpha/m) - F(t)_\alpha]. \qquad (30)$$

Equation (30) agrees with the expression given by Weiner and Ascar [69]. Thence, introducing (6, 27 – 28) into (23 - 25) the quantum equation of the BHO given by Weiner and Forman [66] is obtained.

Moreover, by means of (20, 30) it is possible to write the Hamiltonian bath interaction in the Schrödinger representation as

$$V_e(q, \partial S/\partial q, t) = \beta (S_{(q,t)}/m) - q F(t) + C(t),  \quad (31)$$

or, equivalently

$$V_e(q, \partial S/\partial q, t) = (\beta \hbar /2im) \ln [\psi / \psi^*] - q F(t) + C(t). \quad (32)$$

Where it has been used the definition:

$$\psi_{(q,t)} = A_{(q,t)} \exp[i S_{(q,t)} / \hbar]. \quad (33)$$

By choosing the integrating constant C(t) such as:

$$C(t) = (\beta \hbar /2im) \int_{-\infty}^{+\infty} \psi^* [\ln \psi / \psi^*] \psi \, dx, \quad (34)$$

equation (32) leads to the Schrödinger-Langevin equation given by Kostin [55].

## 3. Discussion and conclusion

It is noteworthy to observe that even if the BHO problem is deterministic-like, the phase space vector $\upsilon^{BHO}_j$ contains a term that is a stochastic process so that actually the time evolution is probabilistic.

This particular case is possible because the bath interaction is described by means of the potential-type expression (30).

The case of the BHO represents a very illuminating example since the existence of the random potential (30) warrants conditions (6, 18) and hence that the QHA has a bi-univocal correspondence with the Schrödinger representation. It must also be noted that the potential (30) exists since the stochastic force in (29) is integrable function of time and moments. From general point of view equations (15.b-15.c) do not always admit a random potential function as for (21). When the noise being stochastic function of the space is considered (i.e., the local character is introduced into the equation of motion) the stochastic potential description is clearly not generally possible as well as the correspondence with the Schrödinger representation.

The physical relevance of the stochastic QHA model goes beyond the reproducing the standard quantum mechanics in the deterministic limit. In fact, it correctly leads to the dissipative Schrödinger equation given by Kostin [55].

Generally speaking, it is noteworthy to observe that is always possible to have the hydrodynamic analog representation (some stochastic cases included) of a problem described by means of the Schrödinger representation but not vice versa. This, since condition (18) is necessary to perform the back transformation [56]. When the environment is considered, the QHA problem can be traced back to the Schrödinger representation if the bath interaction can be described by means of a stochastic potential, as in the case of the Brownian harmonic oscillator, so that condition (18) can be warranted and non-locality is not disrupted.

However, the stochastic potential exists when the stochastic force is an integrable function (as for (8) that is not expressed as a function of space (i.e., non-local)). When the noise is function of space (the local character is introduced into the motion equation), the stochastic potential description is clearly not generally possible and, in this case, the QHA does not have its respective representation in the Schrödinger problem because (18) does not hold anymore.

## Appendix

Let's consider the stochastic differential equation (SDE)

$$\dot{x}_j \, dt = \upsilon^Q_j \, dt + (k \Theta)^{1/2} \mu^{1/2}_{jm} d\mathcal{W}_m, \quad (A.1)$$

so that (2) can be seen as the deterministic limit of a stochastic process where k is the Boltzmann constant, $\Theta$ is the fluctuation amplitude parameter with the dimension of a temperature, $\mu_{mj} = \mu^{1/2}_{mn}\mu^{1/2}_{nj}$ is the associated migration tensor and $d\mathcal{W}_m$ is a white noise with null mean, with unitary variance and $\int d\mathcal{W}_j$ continuous [33].

Given $P(y,x|\tau,t)$, the transition probability function (TPF) that represents the probability that an amount of the PDF at time t, in a temporal interval $\tau$, in a point x, transfers itself from x to a point y [36], by expressing the conservation of the PDF $\rho$ in integral form, it follows that the TPF generates a displacement by a vector $(x,t) - (z,0)$ according to the rule [36]

$$\rho_{(x,t)} = \int_{-\infty}^{+\infty} P(x,z|t,0)\, \rho_{(z,0)}\, d^{6n}z \tag{A.2}$$

and obeys to the Smoluchowski integro-differential equation for Markovian PTF [36]

$$P(x,x_0|t+\tau, t_0) = \int_{-\infty}^{+\infty} P(x,z|\tau, t)\, P(z,x_0|t-t_0, t_0)\, d^{6n}z \tag{A.3}$$

(since (A.3) for $t_0 = 0$, $t = 0$, $\tau = t$ equals (A.2), the TPF is also conserved).

The SDE (A.1) can be integrated by means of the Cauchy-Euler method as follows

$$x_{j\,n+1} = x_{j\,n} + \upsilon^Q_j(x_n,t_n)\Delta t_n + (k\Theta)^{1/2}\mu^{1/2}_{jm}(x_n,t_n)\Delta\mathcal{W}_{m\,n}, \tag{A.4}$$

here

$$x_{j\,n} = x_j(t_n), \tag{A.5}$$

$$\Delta t_n = t_{n+1} - t_n, \tag{A.6}$$

$$\Delta\mathcal{W}_{m\,n} = \mathcal{W}_m(t_{n+1}) - \mathcal{W}_m(t_n), \tag{A.7}$$

where $\Delta\mathcal{W}_m$ has Gaussian zero mean and unitary variance probability function that for $\Delta t_n = \Delta t\ \forall\ n$ reads

$$\mathcal{P}(\Delta\mathcal{W}_m, \Delta t) \propto = [(2\pi\Delta t)^{-3n}]\exp-\tfrac{1}{2}[\Delta\mathcal{W}_m\Delta\mathcal{W}_m/\Delta t]. \tag{A.8}$$

By using (A.4-A.7), $\mathcal{P}(x_{n+1}, x_n|\Delta t, n\Delta t)$ as a function of x reads:

$$\mathcal{P}(x_{n+1},x_n|(n+1)\Delta t, n\Delta t) \propto \exp-\tfrac{1}{2}[(x_{j\,(n+1)} - \langle x_{j(n+1)}\rangle)(k\Theta\Delta t)^{-1}\mu^{-1}_{jm}(x_{m(n+1)} - \langle x_{m(n+1)}\rangle)]$$

$$= \exp-\tfrac{1}{2}[(x_{j\,(n+1)} - (x_{j(n)}+\upsilon^Q_{j(n)}\Delta t))(k\Theta\Delta t)^{-1}\mu^{-1}_{jm}(x_{m(n+1)} - (x_{m(n)} + \upsilon^Q_{m(n)}\Delta t))], \tag{A.9}$$

where $\mu^{-1}_{jm} = (\mu^{1/2}(\mu^{1/2})^T)^{-1}_{jm}$,

where $\upsilon^Q_{(x_n, n\Delta t)}$ (calculated by means of (3,7)) and $\rho_{(x_n, n\Delta t)}$ lead to $\rho_{(x_{n+1}, (n+1)\Delta t)}$ at the following instant by

applying the conservation of the PDF $\rho$ (in the discrete time) that reads

$$\lim_{\Delta t\to 0}\rho_{(x_{n+1}, (n+1)\Delta t)} = \int_{-\infty}^{+\infty} \mathcal{P}(x_{n+1}, x_n|\Delta t, n\Delta t)\, \rho_{(x_n, n\Delta t)}\, d^{6n}x_n. \tag{A.10}$$

from where it follows that $\lim_{\Delta t\to 0} P(x_{n+1}, x_n|\Delta t, n\Delta t) = \mathcal{P}(x_{n+1}, x_n|\Delta t, n\Delta t)$.

For instance, in the case of a diagonal covariance matrix $\mu_{jm} = \mu_{(j)} \delta_{jm}$, the integral path solution [37] for the PDF (A.2) as well as the PTF respectively read

$$\rho(x,t) = \lim_{\Delta t \to 0} \left\{ \prod_{n=0}^{w=(t/\Delta t)-1} \int_{-\infty}^{+\infty} d^{6n}x_n \right\} [(2\pi k\Theta\Delta t)^{-3n}]^{(w+1)/2} \left(\prod_{h=1}^{6n} \mu_{(h)}\right)^{-(w+1)/2}$$

$$\times \exp\left[-\sum_{j=1}^{6n}\sum_{n=0}^{w} \Delta t \left[((x_{n+1}-x_n)/\Delta t) - \upsilon^Q_n - \tfrac{1}{2} d/dt(\upsilon^Q_n)\Delta t\right]^2_j ((k\Theta)^{-1}\mu^{-1}_{(j)})\right] \rho_{(x_0,0)}, \quad (A.11)$$

$$P(x_{(t)}, x_0|t, 0) = \lim_{\Delta t \to 0} \left\{ \prod_{n=1}^{w=(t/\Delta t)-1} \int_{-\infty}^{+\infty} dx_n \right\} [(2\pi k\Theta\Delta t)^{-3n}]^{w/2} \left(\prod_{h=1}^{6n} \mu_{(h)}\right)^{-w/2}$$

$$\times \exp\left[-\sum_{j=1}^{6n}\sum_{n=1}^{w} \Delta t[((x_{n+1}-x_n)/\Delta t) - \upsilon^Q_n - \tfrac{1}{2}d/dt(\upsilon^Q_n)\Delta t]^2_j((k\Theta)^{-1}\mu^{-1}_{(j)})\right] \mathcal{P}(x_1, x_0|\Delta t, 0), \quad (A.12)$$

Where (A.12) obeys to the Smoluchowski equation (A.3).
It is also useful to derive the differential conservation equation for the PTF $P(x,z|t,0)$ that can be obtained by transforming the Smoluchowski equation (A.3) into a differential one by the method of Pontryagin [36] to get:

$$\partial P(x,z|t,0)/\partial t + \partial P(x,z|t,0) \upsilon_j /\partial x_j = 0, \quad (A.13)$$

where the current $P(x,z|t,0) \upsilon_j$ reads

$$P(x,z|t,0) \upsilon_j = P(x,z|t,0) \upsilon^Q_j - \tfrac{1}{2}\partial \, b_{jm} P(x,z|t,0)/\partial x_m + \ldots + (n!)^{-1} \sum_{n=2}^{\infty} \partial^n c^{(n)}_{j\,k\ldots m} P(x,z|t,0)/\partial x_k \ldots \partial x_m, \quad (A.14)$$

with

$$c^{(n)}_{j\,k\ldots h} = \lim_{\tau \to 0} \tau^{-1} \left[\int_{-\infty}^{+\infty} \{(y_j-x_j)(y_k-x_k)\ldots(y_h-x_h)\} P(y,x|\tau,t) \, d^{6n}y_j\right]. \quad (A.15)$$

Usually, in the Hamiltonian (i.e., classical) case, the distance $(y_k-x_k)$ (attainable with a certain probability $P(y,x|\tau,t)$) becomes smaller and smaller as $\tau$ shorter and shorter so that: (1) Velocities $\lim_{\tau \to 0}(y_k-x_k)/\tau$ are finite. (2) The squared mean displacements are proportional to $\tau$ (i.e., $\lim_{\tau \to 0} <(y_k-x_k)^2/\tau> \equiv b_{kk}$ finite). (3) The cumulants $c^{(n)}_{j\,k\ldots h}$ go like $\tau^{n-2}$ and, hence, since $\tau$ can be chosen arbitrarily small, $c^{(n)}_{j\,k\ldots h}$ with $n > 2$ vanishes [36] and the differential equation (A.13) becomes a FPE with Gaussian PTF.
On the contrary, the presence of the quantum potential in the QHA gives a functional dependence of $\upsilon^Q$ by the PDF $\rho_{(x,t)}$, equation (A.13) is not a FPE, the PTF $P(y,x|\tau,t)$ (A.12) is not Gaussian and cumulants higher than two are not null.